\documentstyle[aps,axodraw,epsfig,floats,preprint]{revtex}
\newcommand{\be}{\begin{equation}}
\newcommand{\ee}{\end{equation}}
\newcommand{\bea}{\begin{eqnarray}}
\newcommand{\eea}{\end{eqnarray}}

\def\laq{\ \raise 0.4ex\hbox{$<$}\kern -0.8em\lower 0.62
ex\hbox{$\sim$}\ }
\def\gaq{\ \raise 0.4ex\hbox{$>$}\kern -0.7em\lower 0.62
ex\hbox{$\sim$}\ }
\begin{document}

\title{Candidates for HyperCharge Axion in Extensions of the
Standard Model}

\author{Ram Brustein, David H. Oaknin}
\address{Department of Physics,
Ben-Gurion University, Beer-Sheva 84105, Israel\\
 email: ramyb, doaknin @bgumail.bgu.ac.il}

\maketitle
\begin{abstract}
Many theoretically well-motivated extensions of the Standard
Model contain heavy pseudoscalars that couple to hypercharge
topological density. The cosmological dynamics of such
hypercharge axions could, under certain conditions, lead to
generation of a net baryon number in a sufficient amount to
explain the observed baryon asymmetry in the universe. We examine
the Minimal Supersymmetric Standard Model and string/M-theory
models and determine specific conditions which heavy axion-like
pseudoscalars must satisfy to successfully drive baryogenesis. We
find that all candidates in the Minimal Supersymmetric Standard
Model fail to obey some of the constraints, and that only in
special string/M-theory models some axions may be adequate.
\end{abstract}

\section{Introduction and Summary}

Topologically non-trivial configurations of
hypercharge gauge fields can play a relevant
role in the electroweak~(EW) scenario for
baryogenesis
\cite{Giovannini:1998eg,Giovannini:1998gp,Joyce:1997uy,Thompson:1998qz,Vachaspati:1994ng}.
A hypothetical heavy pseudoscalar
field that couples to hypercharge topological
number density, the hypercharge axion~(HCA), can
exponentially amplify primordial hypermagnetic
fields in the unbroken phase of the EW plasma,
while coherently rolling or oscillating. The
coherent motion provides the three Sakharov's
conditions \cite{Guendelman:1992se} and can
lead, under certain conditions, to  generation
of a net hypercharge topological number that can
survive until the phase transition and then be
converted into a net baryon number in a
sufficient amount to explain the origin of the
baryon asymmetry in the universe
\cite{Brustein:1999du,Brustein:1999rk}.

In \cite{Brustein:1999du,Brustein:1999rk} we
have focused on a simple model with an extra
singlet HCA, $a$, whose only coupling to
Standard Model~(SM) fields is through the
following effective operator
 \be \label{hyper}
{\cal L}_{a YY}= \frac{1}{4 M_Y}\ a {\it
Y}_{\mu\nu} {\widetilde {\it Y}}^{\mu\nu},
 \ee
where ${\it Y}_{\mu\nu}$ is the $U(1)_Y$
hypercharge field strength and ${\widetilde {\it
Y}}^{\mu\nu}$ is its dual. The constant $M_Y$
has units of mass. The HCA is massless at very
high energies and gets a mass $m \sim
\Lambda^2/M_A$  from a generic potential
\be
\label{axionpotential}
 {\widetilde V}(a)=\Lambda^4 V(a/M_A),
 \ee
generated at an energy scale $\Lambda$. The
potential $V$ is a bounded periodic function and
$M_A$ is a normalization mass scale.

Unless some fine-tuning mechanism is effective when the axion
potential is generated, the pseudoscalar is trapped far from the
minimum of its potential and then starts to coherently roll or
oscillate around this minimum until the condensate finally
decays. Typically, the cosmological misalignment of $a$, $\langle
a\rangle_c$, is of the same order of magnitude as $M_A$. The axion
rolls if the Hubble time at the scale of potential generation
$t_H \sim M_p/\Lambda^2$ ($M_p$ being the Planck mass) is shorter
or comparable to the characteristic time $t_s \sim m^{-1}$ for a
coherent oscillation. Otherwise, the axion will oscillate a few
times before the topological condensate decays. While the axion
rolls it generates a net topological number which is stored in
very long wavelength modes, $k \ll \Lambda$, that remains frozen
in the plasma until the EW phase transition. Oscillations should
happen just before or during the EW phase transition so that the
generated topological number, stored in shorter wavelength modes
$k \sim \Lambda$, does not diffuse in the highly conducting
plasma once the oscillations have stopped
\cite{Brustein:1999du,Brustein:1999rk}.

A crucial element in the scenario of  HCA driven
baryogenesis is the requirement that the
misalignment $\langle a\rangle_c \sim M_A$ is
larger than the
decay constant of the HCA into two hypercharge
photons, $M_Y$
\cite{Brustein:1999du,Brustein:1999rk}. If the
axion oscillates, it is enough that $<a>_c$ is
somewhat larger than $M_Y$,
\be
\label{ratio}
<a>_c/M_Y \sim M_A/M_Y > 1,
\ee
but if it rolls, it is necessary that
\be \label{rolls}
<a>_c/M_Y \sim M_A/M_Y > \sqrt{M_P/\Lambda}.
\ee

The heavy HCA differs significantly  from the
original axion proposed by Peccei and Quinn (PQ)
as an elegant solution to the strong CP problem
\cite{Peccei:1977hh,Weinberg:1978ma,Wilczek:1978pj}.
The PQ axion couples to the nonabelian $SU(3)$
topological density
 \be
 \label{QCDaxion} {\cal L}_{a_{PQ} {\it G}{\it G}}=
 \frac{1}{4 M_g}\ a_{PQ} {\it G}_{\mu\nu} {\widetilde
 {\it G}}^{\mu\nu},
 \ee
which generates a potential
\be
\label{QCDpotential}
 {\widetilde V}(a_{PQ})=\Lambda_g^4 V(4\pi^2 a_{PQ}/g^2_s M_g)
 \ee
at the QCD scale $\Lambda_g \sim$~200 MeV through instanton
effects. The QCD axion should get its mass mainly from potential
(\ref{QCDpotential}) in order to fix $\theta_{QCD}=0$. In this
case, the normalization scale for the axion potential $M_A$, is
determined by the axion coupling constant $1/M_g$ such that $M_A
\sim M_g$. This is not the case for the HCA. The axion coupling
(\ref{hyper}) to the abelian hypercharge topological density does
not generate any potential for the HCA, which gets its mass from
some additional sector. In general the scales $M_Y$ and $M_A$ in
the hypercharge sector (\ref{hyper}) and in the mass generation
sector (\ref{axionpotential}) are not related.

In this paper we try to identify  HCA candidates in theoretically
well motivated extensions of the SM. Heavy pseudoscalars that
couple to topological gauge densities appear in low-energy
supersymmetric models with an extended higgs sector. We analyze
here the Minimal Supersymmetric SM (MSSM) as a representative of
this class of models. Numerous HCA candidates appear also in
string/M-theory models. We analyze here 4 dimensional models of
heterotic $E_8 \times E_8$  and Horava-Witten theory as
representatives of this  class of models.

The physical pseudoscalar higgs of the MSSM gets
a mass in the TeV range from soft SUSY breaking
terms or supersymmetric $\mu$-term and couples
to hypercharge topological number density
through anomalous quantum effects when chiral
symmetry is spontaneously broken by the higgs
mechanism. We expect that the scale $M_A$ of the
pseudoscalar potential is about that of SUSY
breaking, $M_A \sim M_{SUSY}$, while the
pseudoscalar inverse coupling to
hypercharge topological density is about the EW
scale, $M_Y \sim M_{EW}$. Typically $M_A$ is
several orders of magnitude larger than $M_Y$,
as required by conditions (\ref{ratio}) and
(\ref{rolls}). As we show this promising picture
is drastically altered at high temperatures in
the unbroken phase of the EW theory. Although
the higgs pseudoscalars get heavy masses from
SUSY breaking terms or $\mu$-term, their
axion-like effective couplings to hypercharge
topological density vanish due to chiral
symmetry restoration\footnote{At this point we
would like to stress that although higgsinos and
gauginos remain massive in the symmetric phase
of the plasma (contrary to leptons and quarks
that become massless), the mass terms they get
from SUSY breaking or $\mu$-term are invariant
under chiral rotations of interaction
eigenstates.}. So the pseudoscalar higgs particles
of the MSSM cannot serve as HCA.

In string/M-theory heavy axions are common. They typically couple
with coupling $1/M_O$ to the topological density of an
``observable" gauge group which is supposed to contain the SM
group, and with coupling $1/M_H$ to the topological density of a
``hidden" gauge group which interacts only through gravitational
strength interactions with the observable sector:
 \be
\label{stringaxion} {\cal L}_{S}=
 \frac{1}{4 M_{O}}\ a {\it F}_{\mu\nu}
 {\widetilde {\it F}}^{\mu\nu} +
 \frac{1}{4 M_H}\ a {\it G}_{\mu\nu}
 {\widetilde {\it G}}^{\mu\nu}.
 \ee
Axions usually get their potentials mainly from
instanton effects in the nonabelian hidden
sector whose field strength is ${\it G}$, while
the abelian hypercharge group $U(1)_Y$ is
contained in the observable sector whose  field
strength is ${\it F}$. In this case, the scale
$M_A$ is given by the inverse axion coupling to
the hidden gauge topological density, $M_H$,
(see eq. (\ref{QCDaxion}) and
(\ref{QCDpotential})), while $M_Y$ is given by
$M_O$.  Conditions (\ref{ratio}) and
(\ref{rolls}), using $M_A/M_Y=M_H/M_O$, require
 that axions couple to the observable sector
more strongly than they do to the hidden sector.
In 4D models  the ratio $M_H/M_O$ is
determined  by the compactification scheme. As
we will show, standard compactifications
restrict parameter space of string candidates
for HCA to regions where it is difficult to
satisfy in a natural way  all the necessary
conditions that allow them to successfully drive
baryogenesis.

\section{HCA in the Minimal Supersymmetric Standard Model.}

The higgs sector of the MSSM contains two
complex $SU(2)_L$ doublets that give masses to
the up and down components of the three families
of SM fermions. The coupling structure of the
MSSM of relevance to us here  appears in each
one of the two higgs Yukawa sectors, and for
each one of the fermion families separately.
Therefore, instead of discussing the full and complicated MSSM
we first discuss, for the sake of simplicity,
a linear $\sigma$-model with a $U(1)_Y$
abelian gauge group, one fermion flavour $\psi$,
and a singlet complex higgs field $\phi$,
\begin{eqnarray}
\label{linear_1}
 {\cal L}_0 = & (-\frac{1}{4}) Y_{\mu \nu} Y^{\mu \nu} + i {\bar \psi}
(\partial_{\mu}+ig'Y_{\mu}) \gamma^{\mu} \psi + (\partial_{\mu}
\phi^*)(\partial^{\mu} \phi) \nonumber \\
           & - (\lambda \phi {\bar \psi} (\frac{1+\gamma^5}{2}) \psi +
h.c.) - \mu^2 (\phi^* \phi - f^2_{PQ})^2,
\end{eqnarray}
which will allow us to analyze relevant issues. We then apply the results
to the full MSSM.

The $\sigma$-model (\ref{linear_1}) is invariant, up to
anomalies, under a global PQ chiral rotation
\begin{equation} \label{rotation}
\psi \rightarrow e^{i \frac{\alpha}{2} \gamma^{5}} \psi \hspace{1.0in}
\phi \rightarrow e^{-i \alpha} \phi,
\end{equation}
which is spontaneously broken by the vacuum
expectation value of the complex scalar field
$<\phi>=f_{PQ}$.

We are interested in the pseudoscalar component
$a$ of the complex higgs that couples axially to
the fermion through the Yukawa sector,
\begin{eqnarray}
\label{extrasector}
{\cal L} = & (-\frac{1}{4}) Y_{\mu \nu} Y^{\mu \nu} + i {\bar \psi}
(\partial_{\mu}+ig'Y_{\mu}) \gamma^{\mu} \psi - m {\bar \psi} \psi
\nonumber \\
           & + (\partial^{\mu} a)^2 + m^2_a a^2 - i \lambda a {\bar \psi}
\gamma^5 \psi.
\end{eqnarray}
Here $m=\lambda f_{PQ}$ is the chiral mass of
the fermion, and $m_a$ is the  pseudoscalar
mass. In the linear $\sigma$-model
(\ref{linear_1}) the chiral symmetry
(\ref{rotation}) protects the pseudoscalar from
getting a perturbative mass, but in the MSSM it
gets a mass in the TeV range from soft
SUSY-breaking terms or supersymmetric
$\mu$-term, and therefore we have introduced it
in this simple model as an extra free parameter.

\begin{figure}
\begin{center}
\begin{picture}(75,120)(0,0)
 \DashLine(-30,60)(20,60){5} \Text(-30,68)[l]{$a$}
 \Text(0,72)[l]{-i$\lambda \gamma_5$}   \Text(45,60)[b]{$\psi$}
 \ArrowLine(21,60)(61,100)
 \ArrowLine(61,99)(61,21)
 \ArrowLine(61,20)(21,60)
 \Photon(62,100)(112,100){3}{4}
 \Text(40,110)[l]{$g' \gamma^{\mu}$}
 \Text(82,115)[l]{$\gamma_Y$}
 \Photon(62,20)(112,20){3}{4}
 \Text(40,10)[l]{$g' \gamma^{\nu}$}
 \Text(82,35)[l]{$\gamma_Y$}
 \Text(135,100)[b]{$k^1_{\mu}(\epsilon^1_{\rho})$}
 \Text(135,20)[b]{$k^2_{\nu}(\epsilon^2_{\sigma})$}
\end{picture}
\caption{\label{triangle}
 HCA decay into two hypercharge photons through a fermion loop.}
\end{center}
\end{figure}
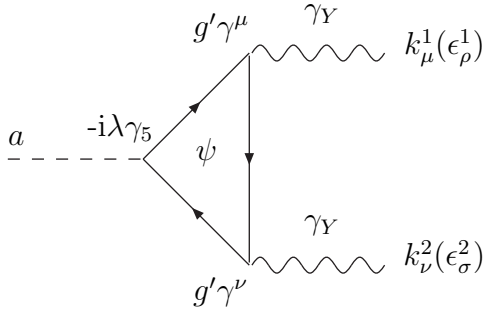

The anomalous coupling of the massive axion to
the  hypercharge topological density
(\ref{hyper}) can be obtained directly in a
standard way from the invariant amplitude for
the decay process of $a$ into two hypercharge
photons $a \rightarrow 2 \gamma_Y$ shown in
Fig.~\ref{triangle},

\begin{eqnarray}
\label{feynmann}
-i {\cal M} &= -2i \lambda g'^2 \int \frac{d^4q}{(2\pi)^4} Tr \left[
\gamma^5
\frac{i}{(\not q - \not k^1 - m)} \gamma^{\mu} \frac{i}{(\not q - m)}
\gamma^{\nu} \frac{i}{(\not q + \not k^2 - m)} \right]
\epsilon^1_{\mu} \epsilon^2_{\nu} \nonumber \\
           &= 8 \lambda g'^2 m
\epsilon^{\mu \nu
\rho \sigma} k^1_{\mu} k^2_{\nu} \epsilon^1_{\rho} \epsilon^2_{\sigma}
\hspace{0.05in} I(k^1,k^2) = \frac{1}{M_Y} \epsilon^{\mu \nu \rho
\sigma} k^1_{\mu} k^2_{\nu} \epsilon^1_{\rho} \epsilon^2_{\sigma},
\end{eqnarray}
and therefore
\be \label{general}
M_Y = \frac{1}{8 \lambda g'^2} \frac{1}{m} |I|^{-1},
\ee
where $k^1$,$k^2$ are the 4-momenta of the two outgoing hypercharge
photons in the center of mass system, and $\epsilon^1$,$\epsilon^2$ are
their polarizations vectors. The scalar function $I$ is given by
\be
\label{intmomentum}
I = \int \frac{d^4q}{(2\pi)^4}
\frac{i}{[(q-k^1)^2-m^2][q^2-m^2][(q+k^2)^2-m^2]}.
\ee
The integral in eq. (\ref{intmomentum}) is finite and can be computed
using standard techniques: \newline
If $0 < \xi=\frac{m_a}{m} \le 2$,
\be
I(\xi) = -\frac{1}{8\pi^2 m^2}
\left[\frac{\makebox{Arcsin}(\xi/2)}{\xi}\right]^2.
\ee
If $2 < \xi=\frac{m_a}{m} < \infty$,
\be
I(\xi) = \left. I \right|_{\xi=2} - \frac{1}{16\pi^2 m^2}
\frac{1}{\xi^2} \left(-\frac{1}{2}\left[ln\left( \frac{1-\tau}{1+\tau}
\right)\right]^2 + i\pi ln(1-\tau^2) \right),
\ee
where $\tau=(1-\frac{4}{\xi^2})^{1/2}$.

The function  $M_Y(\xi)$ reaches its global minimum at $\xi=2$,
when the fermion circulating in the loop is on-shell, \be
\label{maximum} M_Y(\xi=2) = \frac{16}{\lambda g'^2} m =
\frac{16}{g'^2} f_{PQ}. \ee The function decreases slowly from
its value at $\xi=0$ (massless axion) \be \label{massless}
M_Y(\xi=0) = \frac{4\pi^2}{g'^2} f_{PQ}, \ee until it reaches its
minimum and then grows fast towards infinity when $\xi \gg 2$.

The smaller $M_Y$ is, the stronger the axion
coupling (\ref{hyper}) to hypercharge
topological density. If we assume the axion mass
$m_a$ is in the TeV range, we can conclude that
fermions with masses around a few hundreds GeV
or larger contribute significantly to the
anomalous coupling, but much lighter fermions do
not.

At this point we would like to make two comments:
 \newline
 1) The anomalous axion coupling to gauge
topological densities is closely associated with
the spontaneous or explicit breaking of chiral
symmetry in the fermionic mass sector. In the case of a
massless fermion the trace of $\gamma$ matrices
in  (\ref{feynmann}) vanishes, and therefore the
coupling ${\cal M}$ vanishes.
 \newline
 2) The  scale $M_Y$ is fixed by the scale $f_{PQ}$ at
which chiral symmetry is broken.

The effects of high temperature and chiral symmetry restoration
on the anomalous vertex have been studied in the simplest case of
a perturbatively massless axion, $\xi \rightarrow 0$
\cite{Gelis:2000jz,Kumar:2000xm,Gupta:1997gn,Pisarski:1996ne}. We
report the result obtained there for the invariant amplitude:
 \be \label{finite}  -i {\cal M} =
 8 \lambda g'^2 m(T) \epsilon^{\mu \nu \rho
\sigma} k^1_{\mu} k^2_{\nu} \epsilon^1_{\rho}
\epsilon^2_{\sigma} \hspace{0.05in} I_{T},
 \ee
where $m(T)$ is the chiral fermion mass that now depends on the
temperature $T$. At high temperature the scalar function $I_{T}$
is proportional to $\frac{1}{x(T) T}$, where $x(T)$ is the
infrared cut-off. Far below the phase transition, the infrared
cut-off is provided by the mass of the fermion $x(T) \sim m(T)$;
therefore the invariant amplitude depends on $T$ as $\frac{1}{T}$,
and its dependence on the fermion mass cancels. At temperatures
close to chiral symmetry restoration, $m(T) \rightarrow 0$,
hard-thermal corrections are important and provide the effective
cut-off, $x(T) \sim g' T$. As a consequence, the anomalous
invariant amplitude goes like $\frac{m(T)}{g' T^2}$ and strictly
vanishes in the symmetric phase of the plasma when $m(T)=0$. This
conclusion can be generalized to the case of a massive axion,
following the same argument: the axion mass provides an
additional infrared cut-off, $x(T) \sim m_a$, but the anomalous
coupling still vanishes because of chiral symmetry restoration.

The analysis we have presented can be immediately extended to the
specific setup of the MSSM. The MSSM contains two higgs
pseudoscalars that couple independently to hypercharge
topological density when the EW symmetry is spontaneously broken
by the non-zero higgs vacuum expectation values, $f^1_{PQ},
f^2_{PQ} \sim M_{EW}=100$~GeV. In the broken phase of the plasma
one of the two pseudoscalar mass eigenstates remains massless, up
to anomalies: it is the pseudogoldstone boson that is absorbed as
the longitudinal component of the massive Z. The second mass
eigenstate gets a mass typically in the TeV range from soft SUSY
breaking terms or $\mu$-term and, therefore, it could be a
possible candidate for HCA. The largest contribution from SM
fermions to the heavy pseudoscalar coupling to hypercharge
topological density comes from the heavy top quark
(\ref{maximum}), \be \label{MSSMbroken} M_Y \sim \frac{1}{3}
\times \frac{16 M_{EW}}{g'^2} \sim 50~TeV. \ee The factor
$\frac{1}{3}$ takes into account three different colors for the
quark.

The heavy pseudoscalar higgs has additional axial coupling to
charginos and neutralinos, through three-point vertices
higgs-higgsino-gaugino. These are supersymmetric vertices that
are not generated through the spontaneous breaking of the EW
symmetry. The dimensionless coupling $\lambda$ in each
sector is fixed by SUSY: $\lambda=g/\sqrt{2}$ for W-inos and
$\lambda=g'/\sqrt{2}$ for Hyper-inos. These vertices generate at
1-loop an additional contribution to the anomalous pseudoscalar
coupling (\ref{hyper}), $M_Y \sim \frac{16 m_f}{\lambda g'^2}$,
of the order of (\ref{MSSMbroken}), if charginos or neutralinos
masses $m_f$ are not much heavier than the top mass.

The picture described above changes drastically at temperatures
above the EW phase transition. Both pseudoscalar mass eigenstates
are massive but, as we discussed above eq.~(\ref{finite}), the
contribution from SM fermions to their anomalous coupling to
topological densities vanishes because the fermion chiral masses
vanish.

Charginos and neutralinos, on the other hand,
remain massive in the symmetric phase and their
contributions could be nonvanishing. We will
describe here the chargino sector
\cite{Kuroda:1999ks}, but the situation is very
similar for neutralinos. There are two charginos
which contain four Weyl spinors,
$\lambda^+,\lambda^-$ (W-inos), ${\widetilde
H}_1^-,{\widetilde H}_2^+$ (charged higgsinos).
The mass term in this sector in the symmetric
phase of the MSSM is:
\be \label{charginos}
\left(
\begin{array}{l}
 \lambda^+ \\
 {\widetilde H}_2^+
\end{array}
\right)^t
\left(
\begin{array}{l}
 M_2~~~~0 \\
  0~~~~~~\mu
\end{array}
\right)
\left(
\begin{array}{l}
 \lambda^- \\
 {\widetilde H}_1^-
\end{array}
\right) + h.c. =
\mu {\bar \chi_1} \chi_1 + M_2 {\bar \chi_2} \chi_2,
\ee
where we have defined mass eigenstates Dirac spinors,
$
\chi_1 =
\left(
\begin{array}{l}
 {\widetilde H}_1^- \\
 {\overline {\widetilde H}_2^+}
\end{array}
\right),
$
and
$
\chi_2 =
\left(
\begin{array}{l}
 \lambda^- \\
 {\overline \lambda^+}
\end{array}
\right).
$
The parameters $\mu$ and $M_2$ are a higgs-higgs coupling constant and a
soft SUSY breaking parameter, respectively. The mass term
(\ref{charginos}) is invariant under a chiral rotation with opposite
charges of the Dirac interactions eigenstates,
$
\psi_1 =
\left(
\begin{array}{l}
 {\widetilde H}_1^- \\
 {\overline \lambda^+}
\end{array}
\right),
$
and
$
\psi_2 =
\left(
\begin{array}{l}
 \lambda^- \\
 {\overline {\widetilde H}_2^+}
\end{array}
\right),
$
and we can guess that the anomalous coupling
will vanish because of chiral symmetry
restoration. In fact, this is what happens: if
we write the pseudoscalar-higgsino-Wino axial
vertex,
\be
{\cal L}_{axial} = \frac{ig}{\sqrt{2}}\left({\widetilde H}_2^+ \lambda^-
a_2
- \lambda^+ {\widetilde H}_1^- a_1 \right),
\ee
where $a_1$,$a_2$ are the two massive pseudoscalars, in terms of chargino
chiral mass eigenstates in the symmetric phase of the plasma
$\chi_1$,$\chi_2$
\be
{\cal L}_{axial} =
\frac{ig}{\sqrt{2}}\left({\bar \chi_1}
\frac{1+\gamma^5}{2} \chi_2 a_2 - {\bar \chi_2}
\frac{1+\gamma^5}{2} \chi_1 a_1 \right) + h.c.,
\ee we obtain only non-diagonal $\chi_1 -
\chi_2$ vertices, while the
hyperphoton-chargino-chargino only vertex,
$Y_{\mu} {\bar \chi_1} \gamma^{\mu} \chi_1$, is
diagonal. The loop shown in Fig. \ref{triangle}
cannot be closed. Therefore, although charginos
and neutralinos are massive in the symmetric
phase of the plasma, their contributions to the
anomalous 1-loop coupling of pseudoscalars to
hypercharge topological density also vanish
because of chiral symmetry restoration.

The mass matrix (\ref{charginos}) gets thermal contributions that
we have not written explicitly. It is not clear if non-diagonal
thermal terms appear in the mass matrix but, in any case, they do
not break the restored chiral symmetry and our conclusion is not
expected to be modified: the MSSM extended higgs sector does not
contain an HCA capable to drive baryogenesis while coherently
rolling or oscillating.

The mechanism described in \cite{Brustein:1999du,Brustein:1999rk}
for HCA-driven baryogenesis through coherent amplification of
hypercharge fields in the symmetric phase of the EW plasma could
possibly serve to amplify primordial ordinary electromagnetic
fields in the broken phase of the plasma \cite{Tornkvist:2000ay}.
From our previous analysis, it may seem that the heavy
pseudoscalar of the MSSM could have a relevant role: if, as it is
generally assumed, the SUSY breaking terms that give the
pseudoscalar its mass are generated at an intermediate scale
$\sim 10^8 GeV$, the coherent motion of the pseudoscalar when its
potential is generated could have a typical misalignment $\langle
a\rangle_c$, which is several orders of magnitude larger than the
inverse coupling $M_Y \sim 50$~TeV (see (\ref{MSSMbroken})), as
required by conditions (\ref{ratio}) and (\ref{rolls}) for
coherent amplification of the magnetic modes.

But, if the potential is generated at the
intermediate SUSY breaking scale and gives the
pseudoscalar a mass in the TeV range, the
typical time for a coherent oscillation is much
shorter than the Hubble time at the epoch of
potential generation. In
\cite{Brustein:1999du,Brustein:1999rk} we have
considered a singlet HCA that couples only to
hypercharge photons through the operator
(\ref{hyper}). The heavy pseudoscalar of the
MSSM, on the contrary, couples to many other
fields so that the topological condensate decays
incoherently after a few oscillations at
temperatures much above the electroweak phase
transition and does not survive until the EW
phase transition when the coupling (\ref{hyper}) is
generated. So our conclusion is that the
pseudoscalar higgs of the MSSM cannot amplify
ordinary electromagnetic fields.

The MSSM contains additional pseudoscalar
neutral fields, {\it i.e.} pseudoscalar components of
sneutrinos, which
could, in principle, play a similar role to HCA
in amplification of hypercharge topological
number. But for sneutrinos, a dimension five
operator similar to (\ref{hyper}) is forbidden
by R-parity symmetry. Extensions of the MSSM
without R-parity in which sneutrinos couple to
topological gauge densities have been discussed
in \cite{Bar-Shalom:1999xz} but, as in the case
of the pseudoscalar higgs, the coupling to
hypercharge topological number is generated
through chiral symmetry breaking and therefore
it also vanishes in the unbroken phase of the
plasma.

\section{HCA in string/M-theory models}

In this section we consider possible candidates for HCA in low
energy 4D effective field theory of weakly coupled heterotic $E_8
\times E_8$ string theory (HE) and Horava-Witten (HW) theory
\cite{Horava:1996qa,Horava:1996vs}.
Axions in such theories were studied in detail in
\cite{Witten:1984dg,Choi:1985bz,Banks:1996ss,Choi:1997an}
 and we recall here the necessary results to be able
to discuss their relevance and application to the subject at hand:
HCA's.

Compactified string/M-theory models invariably contain a
model-independent axion $a_{MI}$. In string theory compactifications
$a_{MI}$ corresponds to a perturbatively massless pseudoscalar
mode of the antisymmetric 2-index tensor $B_{\mu\nu}$, while in
HW theory compactifications $a_{MI}$ corresponds to a
perturbatively massless mode of the 3-index antisymmetric tensor
$C_{\mu\nu\rho}$. In compactifications preserving at least $N=1$ 4D
supergravity, such as Calabi-Yau compactifications of HE
string theory or HW theory, additional model-dependent axions $a^i_{MD}$
exist: $a_{MI}$ is the pseudoscalar component of the dilaton superfield
and $a^i_{MD}$ are the pseudoscalar component of moduli superfields.

String/M-theory compactifications contain the SM fields as part of
the so-called ``observable" sector, and additional matter and
gauge fields which couple to the observable sector only through
gravitational strength interactions in the so-called ``hidden"
sector. Stringy axions generically couple to gauge topological
density of both sectors. We will be interested, of course, in the
linear combination of all the axions which couples to observable
hypercharge topological density. Orthogonal linear combinations
couple only to the hidden sector density or do not couple to
gauge topological densities at all. The axion that is of interest
to us couples with coupling $1/M_O$ to the density of the
observable group, whose field strength is ${\it F}$, and with
coupling $1/M_H$ to the density of the hidden group whose field
strength is ${\it G}$. This perturvatively massless axion usually
gets its potential mainly from instanton effects in the
nonabelian hidden sector, while the abelian hypercharge group
$U(1)_Y$ is contained in the observable sector. In this case, the
scale $M_A$ is given by $M_H$, as in eq. (\ref{QCDaxion}) and
(\ref{QCDpotential}), while $M_Y$ is given by $M_O$. Conditions
(\ref{ratio}) and (\ref{rolls}), using $M_A/M_Y=M_H/M_O$, require
that the axion couples to the observable sector more strongly
than it does to the hidden sector. In 4D models the ratio
$M_H/M_O$ is determined  by the compactification scheme.

The model-independent axion $a_{MI}$ coupling to topological gauge
density in HE theory
 \be
 \label{MI} {\cal L}_{MI} =
\frac{1}{4 M_1} a_{MI} ({\it F}_{\mu \nu} {\widetilde {\it
F}}^{\mu \nu} + {\it G}_{\mu \nu} {\widetilde {\it G}}^{\mu \nu}),
\ee
 can be obtained directly from the Bianchi identity of the
3-index antisymmetric tensor, $dH =-tr F^2 - tr G^2 + tr R^2$.
The normalization mass scale $M_1$ is determined by the ratio
between the Yang-Mills coupling constant and the gravitational
constant, and it does not depend on the details of the
compactification scheme. In weakly coupled HE theory $M_1 \sim 7
\times 10^{15} GeV$. Similarly, in  HW theory the coupling of
$a_{MI}$ can be obtained from the Bianchi identity of the 4-index
antisymmetric field strength tensor. The mass scale $M_1$ is not
determined as well as it is in weakly coupled HE theory since it
also depends on the length of the eleventh dimension interval,
and on some additional theoretical input, but it is typically
somewhat higher, $\sim 10^{17} GeV$.

The model-dependent axions $a^i_{MD}$ coupling to topological
gauge density in HE theory is contained in the 10D Wess-Zumino
term $S_{WZ} = c \int d^{10}x (B tr {\it F}^4 + B tr {\it G}^{4} +
...)$ necessary to cancel the anomalies of the theory,
 \bea
  \label{modeldependent} {\cal L}_{MD} &=
\sum_i\frac{1}{4 M^i_2} a^i_{MD} ({\it F}_{\mu \nu} {\widetilde
{\it F}}^{\mu \nu} - l^i {\it G}_{\mu \nu} {\widetilde {\it
G}}^{\mu \nu}).
 \eea
We will focus on the linear combination of model-dependent
axions, the overall model-dependent axion $a_{MD}=M_2(\Sigma_i
a^i/M^i_2)$, where $M_2=(\Sigma_i (1/M^i_2)^2)^{-1/2}$, that
couples to the observable gauge density. Then,
 \be
             {\cal L}_{MD} = \frac{1}{4 M_2} a_{MD}
({\it F}_{\mu \nu} {\widetilde {\it F}}^{\mu \nu}
- l {\it G}_{\mu \nu} {\widetilde {\it G}}^{\mu \nu}).
 \ee
The normalization mass scale $M_2$ is determined by the detailed
properties of the compactification scheme. In HW
compactifications the coupling of $a_{MD}$ can be obtained by
treating HW theory as a strongly coupled HE theory, and varying
the coupling continuously from weak to strong coupling. The
argument is that the axion coupling is determined by some
topological invariants and does not change in the process.
Typically, $M_2 \laq M_1$. In HE models we are considering
$l^i=+1$, and therefore $l=+1$, is required by the fact that the
three-index antisymmetric tensor $H$ is globally defined, $\int
dH=0$. In HW models $l=+1$ is also required for similar reasons
\cite{Choi:1997an}, but we keep $l$ here as a free parameter for
the sake of generality.

The mass generation scale $\Lambda$ in string/M-theory models
depends on hidden sector matter content and interactions in
addition to its dependence on compactification details. It is
therefore not so well determined, and can be anywhere below
$M_p$. But in many models, such as gaugino condensation models it
is expected to be of order $\Lambda^2\sim m_{3/2} M_p$, where
$m_{3/2}$ is the gravitino mass, so $\Lambda\sim 10^{11} GeV$.
However, in \cite{Choi:1997an} it is argued that in some cases
$\Lambda$ can be much smaller. Our attitude here is to take
$\Lambda$ as a free parameter, and find out the constraints on it.

We have collected all relevant information about string/M-theory
axions and turn to discuss implications and consequences for HCA.
The HCA, $a$, is the linear combination of $a_{MI}$ and $a_{MD}$
that couples to the gauge observable sector topological density:
 \be
 a = M_O\left[\frac{1}{M_1}a_{MI}+\frac{1}{M_2}a_{MD}\right],
 \ee
where
 \be
 \label{mo}
 M_O=[M_1^{-2}+M_2^{-2}]^{-1/2}.
 \ee
The HCA coupling to the hidden sector topological density is given
by $1/M_H$, where
 \be
 \label{mh}
  M_H = M_O \frac{M_1^2+M_2^2}{|M_2^2-lM_1^2|}.
  \ee

Condition (\ref{rolls}) for successful baryogenesis driven by $a$
while it is rolling requires that $M_H/M_O>
\sqrt{\frac{M_p}{\Lambda}}$. Using (\ref{mh}), (\ref{mo}),
condition (\ref{rolls}) then becomes a condition on mass scales
$M_1$, $M_2$ and $\Lambda$,
 \be
 \label{tuning}
 |M_2^2 - l M_1^2| \laq \sqrt{\frac{\Lambda}{M_P}} (M_1^2+M_2^2),
 \ee
that cannot be naturally satisfied in standard
compactification schemes, in which $\Lambda\ll M_p$ and $l=1$, unless
scales $M_1$ and $M_2$ are appropriately tuned by some mechanism.
The condition could be more easily realized in special models
in which either $l\neq 1$, or $\Lambda\sim M_p$.

Condition (\ref{ratio}) for baryogenesis induced by an
oscillating axion, on the other hand, is naturally satisfied
since  it requires that $M_H/M_O \gaq 1$. For models in which
$l=1$ this is obviously satisfied since
$\frac{M_1^2+M_2^2}{|M_2^2 - M_1^2|}> 1$. But, as explained in
the introduction, to successfully drive baryogenesis, HCA
oscillations have to occur just before the EW phase transition.
Therefore, the HCA potential has to be generated at temperatures
close to the EW phase transition: $\Lambda \sim M_{EW}$. Additionally,
the HCA mass  $m_a \sim
\frac{\Lambda^2}{M_H}$, has to be about the EW scale,
$m_a \sim M_{EW}$. It follows that the HCA
coupling to the hidden sector has to satisfy $M_H \sim M_{EW}$,
which  can be fulfilled only if at least one of the mass scales
$M_1$, $M_2$ is about the EW scale. This is not expected in
typical string/M-theory models, but may perhaps be viable in
models with a very low string scale \cite{Dienes:1999gw}, or in
models in which HCA is protected by some symmetry. In this range
of parameters, $M_Y \laq M_{EW}$ and the HCA would be directly
detectable in future colliders through its decay into 2 photons
\cite{Brustein:2000it}.

\section{Conclusions.}

We have studied the MSSM and 4D string/M-theory low energy
effective field theories, looking for suitable candidates for HCA.

The natural candidate in the MSSM is the heavy higgs pseudoscalar
that couples in the broken phase of EW theory to topological gauge
densities through 1-loop triangle diagrams. But, as we have shown
in section II, this coupling vanishes in the symmetric phase of
the EW theory due to chiral symmetry restoration in the fermionic
mass sector of the theory. The coupling of other possible
candidates, {\it i.e.} pseudoscalar components of sneutrinos, to
topological gauge densities is forbidden by R-parity symmetry. We
conclude that the MSSM does not contain an HCA that can
successfully drive baryogenesis. In some models with broken
R-parity, sneutrinos do couple to topological gauge densities
through triangle diagrams but also fail to serve as HCA's due to
chiral symmetry restoration in the unbroken phase of the EW
theory.

Stringy axions couple directly to hypercharge topological density
at the compactification scale, which is typically much higher than
the EW scale. But, we have concluded in section III that in
generic compactifications the specific conditions for successful
baryogenesis are violated in one way or another. We have outlined
requirements for more elaborate models which may lead to HCA's
which obey all the conditions. Such models may be realized in
some special compactification schemes, or in scenarios of very low
string scale.

\acknowledgments
We want to gratefully acknowledge stimulating and very
useful comments by J.R.~Espinosa about the symmetric phase of the MSSM.

\end{document}